\def\cm2{cm$^{-2}$}
\def\nhi{\mbox{$N_{\rm HI}$}}
\def\logh{\mbox{$\log (N_{\rm HI}$)}}

\def\e{et~al.}

\def\approxlt{\mathrel{\spose{\lower 3pt\hbox{$\sim$}}
        \raise 2.0pt\hbox{$<$}}}
\def\approxgt{\mathrel{\spose{\lower 3pt\hbox{$\sim$}}
        \raise 2.0pt\hbox{$>$}}}

\def\h{{\rm H}\,{\sc I}}
\def\f{{\rm Fe}\,{\sc II}}
\def\m{{\rm Mg}\,{\sc I}}
\def\M{{\rm Mg}\,{\sc II}}

\def\c2{{\rm C}\,{\sc II}}
\def\c4{{\rm C}\,{\sc IV}}
\def\s2{{\rm Si}\,{\sc II}}
\def\s4{{\rm Si}\,{\sc IV}}
\def\a2{{\rm Al}\,{\sc II}}
\def\a3{{\rm Al}\,{\sc III}}
\def\o1{{\rm O}\,{\sc I}}
\def\n1{{\rm N}\,{\sc I}}

\documentclass[useAMS]{mn2e}

\usepackage{supertabular}
\usepackage{subfigure}
\usepackage{lscape}
\usepackage{longtable}
\usepackage{psfig}
\usepackage{rotating}

\usepackage{graphicx}

\title[Classical and MgII-selected DLAs: impact
on $\Omega_{\rm HI}$ at $z<1.7$.]
{Classical and MgII-selected Damped Lyman-$\alpha$ Absorbers: impact
on $\Omega_{\rm HI}$ at $z<1.7$.}

\author[C\'eline P\'eroux et al.]
        {C\'eline P\'eroux$^{1}$\thanks{Marie Curie Fellow}\thanks{e-mail: cperoux@eso.org},
        Jean-Michel Deharveng$^{2}$, Vincent Le Brun$^{2}$ 
\newauthor
\& Stefano Cristiani$^{1}$ 
\\
$1$ Osservatorio Astronomico di Trieste, Via G. B. Tiepolo 11, 34131 Trieste, Italy.\\
$2$ Laboratoire d'Astrophysique de Marseille, Traverse du Siphon, Les trois Lucs
BP 8, 13376 Marseille Cedex 12, France.\\
}

\begin{document}

\date{Accepted. Received}

\pagerange{\pageref{firstpage}--\pageref{lastpage}} \pubyear{2004}

\maketitle

\label{firstpage}
%

\begin{abstract}

The Damped-Ly$\alpha$ systems (DLAs), seen in absorption in the
spectrum of background quasars, are believed to contain a large
fraction of the neutral gas in the Universe. Paradoxically, these
systems are more difficult to observe at $z_{\rm abs}<1.7$, since they
are rare and their \h\ feature then falls in UV spectra, thus
requiring the use of space-borne facilities. In order to overcome this
observational difficulty, Rao $\&$ Turnshek (2000) pioneered a method
based on
\M-selected DLAs, that is absorbers discovered thanks to our knowledge
of their \M\ feature in optical spectra. In the present work, we use
new observations undertaken at the TNG as well as a careful literature
\& archival search to build samples of low redshift absorbers
classified according to the technique used for their discovery.  We
successfully recover \nhi\ and equivalent widths of \f\ 2600, \M\
2796, \M\ 2803 and \m\ 2852 for a sample of 36 absorbers, 21 of which
are \M-selected. We find that the \M-selected sample contains a
marginally larger fraction of absorbers with \logh$>$21.0 than seen
otherwise at low redshift. If confirmed, this property will in turn
affect estimates of $\Omega_{\rm HI}$ which is dominated by the
highest \h\ column densities. We investigate on the source of the
potential discrepancy and find that \logh\ does {\it not} correlate
significantly with metal equivalent widths. Similarly, we find no
evidence that gravitational lensing, the fraction of associated
systems or redshift evolution affect the absorber samples in a
different way. We conclude that the hint of discrepancies in
\nhi\ distributions between the \M-selected DLAs and the others most
likely arises from small number statistics. Therefore, further
observations based on both \h\ and \M\ selection techniques are
required to better clarify the impact on estimates of $\Omega_{\rm
HI}$ at low redshift.

\end{abstract}
%

%

\begin{table*}
\begin{center}
\caption{Journal of the TNG observations for metal equivalent width measurements.}
\begin{tabular}{lllccc}
\hline
Coordinate & Alternative   &$z_{\rm em}$   &Mag &Obs Date    &Exp Time\\
Name &  Name   &    &  &    &(sec)\\
\hline
Q0217$+$0144$^a$		&0215$+$015       &1.715   &18.3         &19 Feb 2003 &1800$\times$2\\	
Q0456$+$0400  			&Q0454$+$039      &1.345   &16.5         &18 Feb 2003 &1800$\times$2\\	
Q0813$+$4813  			&3C196            &0.871   &17.8         &18 Feb 2003 &1800$\times$2\\	
Q0938$+$4129  			&Q0935$+$417      &1.980   &16.2         &18 Feb 2003 &1800$\times$2\\		
Q1001$+$5553  			&Q0957$+$561      &1.414   &16.7         &19 Feb 2003 &1800$\times$2\\	
Q1124$-$1705  			&HE1122$-$1649    &2.400   &16.5         &19 Feb 2003 &1800$\times$2\\	
Q1211$+$1030    		&Q1209$+$107  	  &2.187   &17.6         &19 Feb 2003 &1800$\times$2\\	
Q1232$-$0224    		&PKS1229$-$021	  &1.045   &17.6         &21 Feb 2003 &1800$\times$2\\	
Q1250$+$2631  			&Q1247$+$267      &2.038   &15.8         &19 Feb 2003 &1800$\times$1\\
...				&...		  &...	   &...	         &21 Feb 2003 &900$\times$2 \\		
Q1331$+$3030  			&Q1328$+$307      &0.849   &17.2         &22 Mar 2003 &1800$\times$2\\
Q1354$+$3139  			&Q1351$+$318      &1.326   &17.4         &26 Mar 2003 &1800$\times$2\\	
Q1624$+$2345  			&Q1622$+$238      &0.927   &17.5         &10 Apr 2003 &1800$\times$2\\	
Q1631$+$1156  			&Q1629$+$120      &1.795   &18.5         &10 Apr 2003 &1800$\times$1\\
...				&...		  &... 	   &...		 &11 Apr 2003 &1800$\times$1\\		
\hline
\end{tabular}

\vspace{0.1cm}

$^a$ Quasar with Broad Absorption Lines (BAL).\\
\end{center}
\label{t:JoO}
\end{table*}

\section{Introduction}

Intervening absorption systems in the line of sight towards distant
quasars allow for direct observation of the distribution of gaseous
matter from the epoch of initial galaxy formation to the present day
(Lanzetta \e\ 1991; Wolfe \e\ 1995). In particular, these absorbers
provide important ways of measuring the neutral gas and metallicity
content of the Universe at high-redshifts (Vladilo \e\ 2000; Savaglio
2001; Kulkarni \& Fall 2002; Prochaska \e\ 2003; P\'eroux \e\
2003a). The high-column density end of this population is composed of
the Damped-Ly$\alpha$ systems (DLAs) which have \h\ column densities,
\logh, $>$20.3 cm$^{-2}$ and the sub-Damped-Ly$\alpha$ systems
(sub-DLAs) with $\log$ 19.0$<$N(HI)$<$20.3 cm$^{-2}$. These two are found
to contain a large fraction of the neutral gas in the Universe
(P\'eroux et~al. 2003b). Therefore, these systems can be used to
compute the ratio of \h\ density to the critical density of the
Universe, $\Omega_b$, which in turn provides information on the gas
consumption and star formation over time.

This cosmological mass density is rather well constrained
at high-redshift (Storrie-Lombardi et~al. 1996; Storrie-Lombardi \&
Wolfe 2000; P\'eroux et~al. 2003b). In contrast, measurements of
$\Omega_{\rm HI}$ at $z < 1.7$ are paradoxically more difficult for
several reasons: the observed absorber wavelengths are shifted to the
ultraviolet requiring space observations and the geometry of the
Universe combined with the paucity of DLA systems requires the
observations of many quasar lines of sight. Nevertheless, this
redshift range is of particular importance since it comprises
$\sim$70\% of the look-back time ($\Omega_{\rm M}=0.3; \Omega_{\rm
\Lambda}=0.7$ cosmology). 

Lanzetta et~al. (1995) were the first to derive $\Omega_{\rm HI}$ at
low redshift using the International Ultraviolet Explorer (IUE)
satellite. Rao, Turnshek \& Briggs (1995) proposed a new method based
on the observational evidence which indicates that DLAs are always
associated with a \M\ system, while the reverse is not true. From
successful HST observations of a sample of \M-selected systems (Rao \&
Turnshek 2000), they estimate DLAs statistics by correcting for the
observationally known incidence of \M\ systems in a random quasar
sample. They then compute the mass of neutral gas using their
``derived'' sample of low redshift DLAs. Using the same technique,
Churchill (2001) has also discovered new DLAs at very low redshifts:
$z \sim 0.05$. These observations imply a rather flat distribution of
the cosmological mass density of neutral hydrogen down to low
redshift, at odds with current ideas and models of cosmic star
formation (Pei \& Fall 1995; Nagamine, Springel \& Hernquist
2003). Moreover, local estimates of \h\ mass measured by Zwaan et
al. (1997), and recently confirmed by Zwaan et al. (2003) are
difficult to reconcile with such high values of $\Omega_{\rm HI}$ in
low redshift quasar absorbers.

In the present work, we compare the observed properties of
\M-selected absorbers  with more ``classical'' DLAs/sub-DLAs. 
The goal is to identify whether both observational methods sample the
column density parameter space and do not impact the calculation of
the cosmological mass density at low redshift. In section 2, we
present the data samples which are from both a targeted observational
program and a dedicated literature and archival search. We describe
the data reduction process as well as the method used to make the
measurements of \logh\ and metal equivalent widths (hereafter
EW). Constructing well-defined samples is challenging since the data
have been taken at various epochs by different authors having distinct
scientific goals in mind. Nevertheless, we have been extremely
conservative in selecting the systems and, for the purpose of our
analysis, we have also split the non-\M\ selected samples into two
sub-samples which allow for comparison of the various discovery
techniques. Details on this matter are given in sections 3 and 4 where
we also compare the properties of the samples of classical and
\M-selected quasar absorbers.

\begin{figure*}
   \includegraphics[width=70mm]{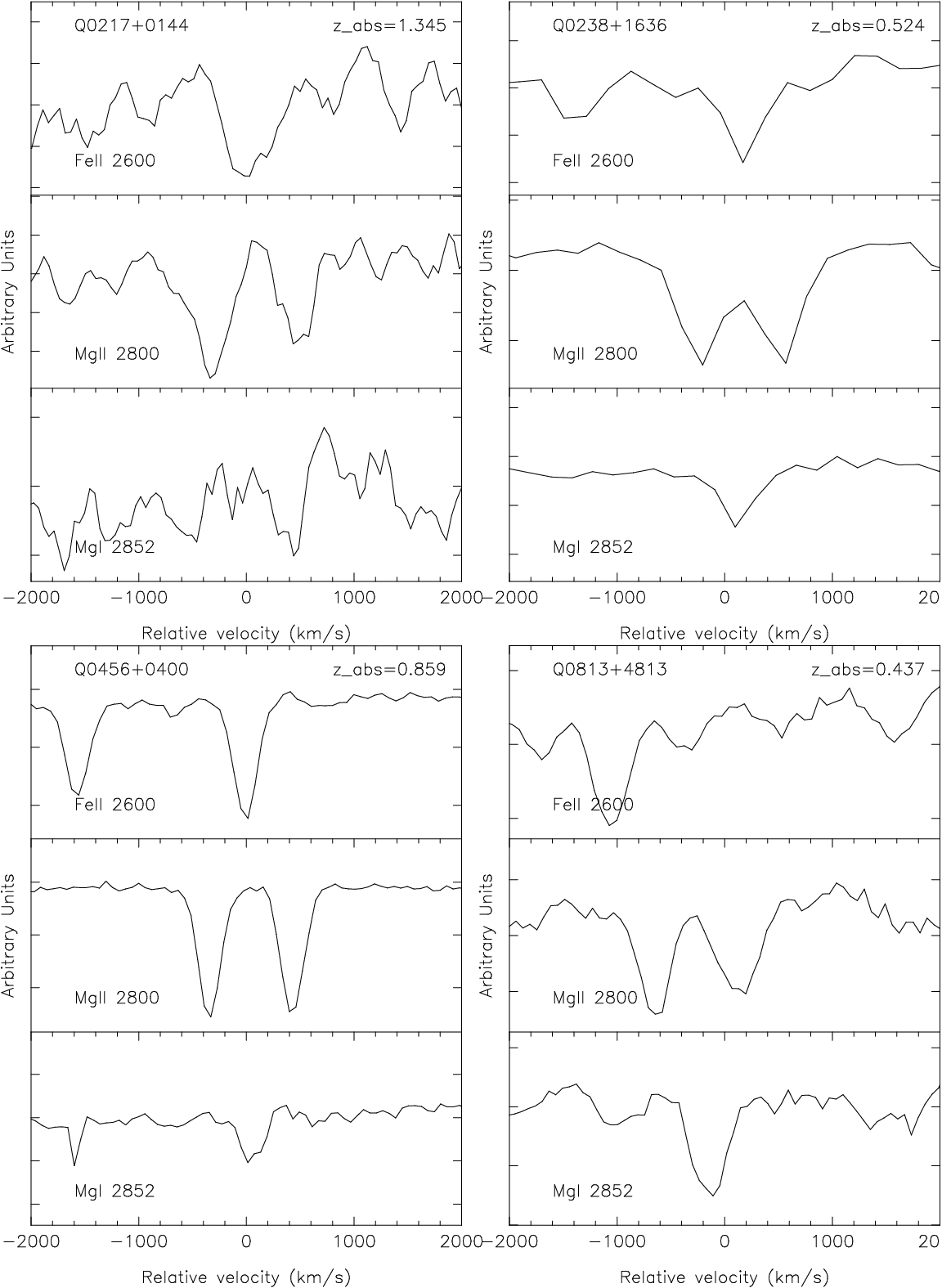}
\vspace{1.cm}		  
   \includegraphics[width=70mm]{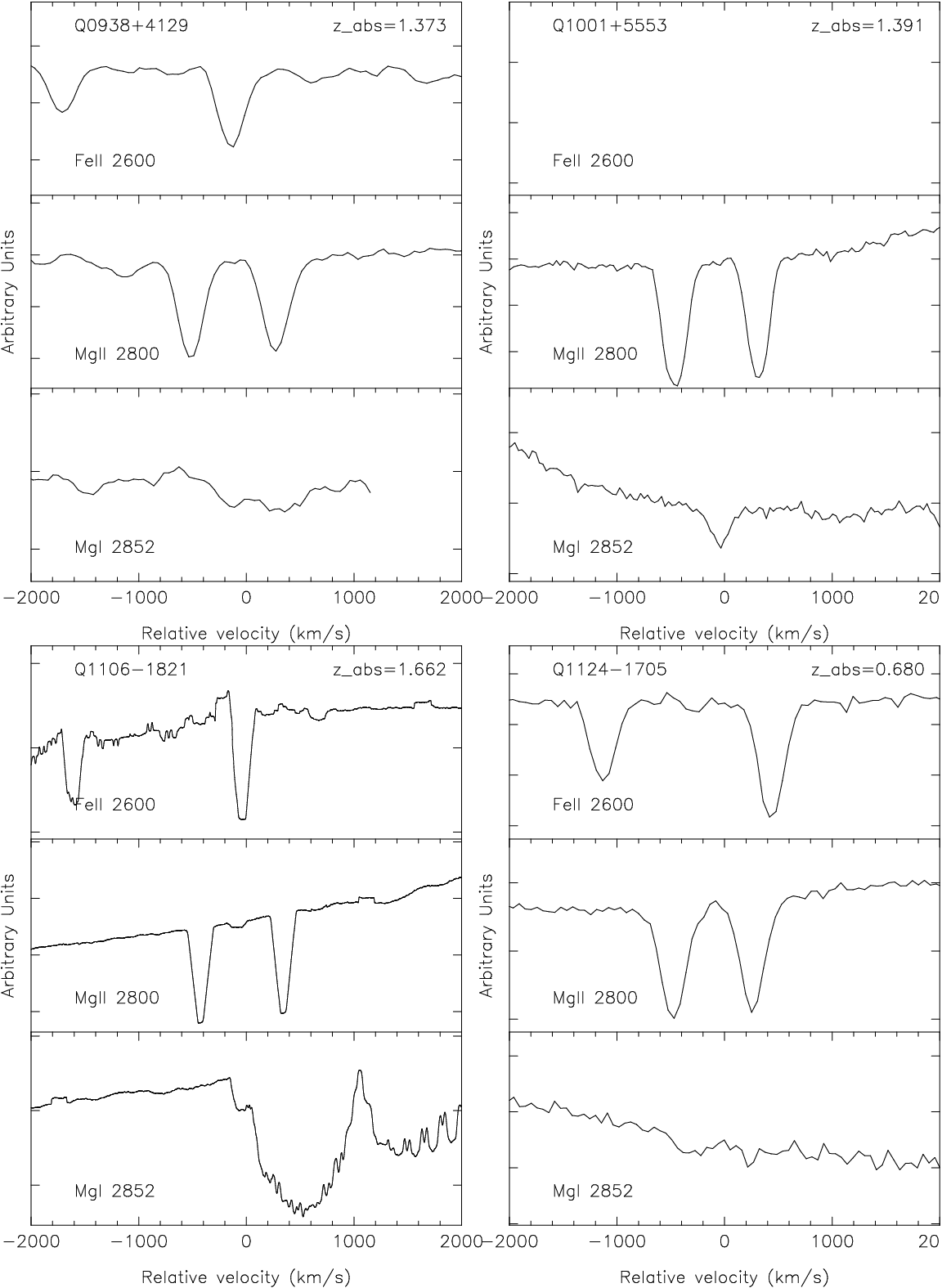}
\vspace{1.cm}		  
   \includegraphics[width=70mm]{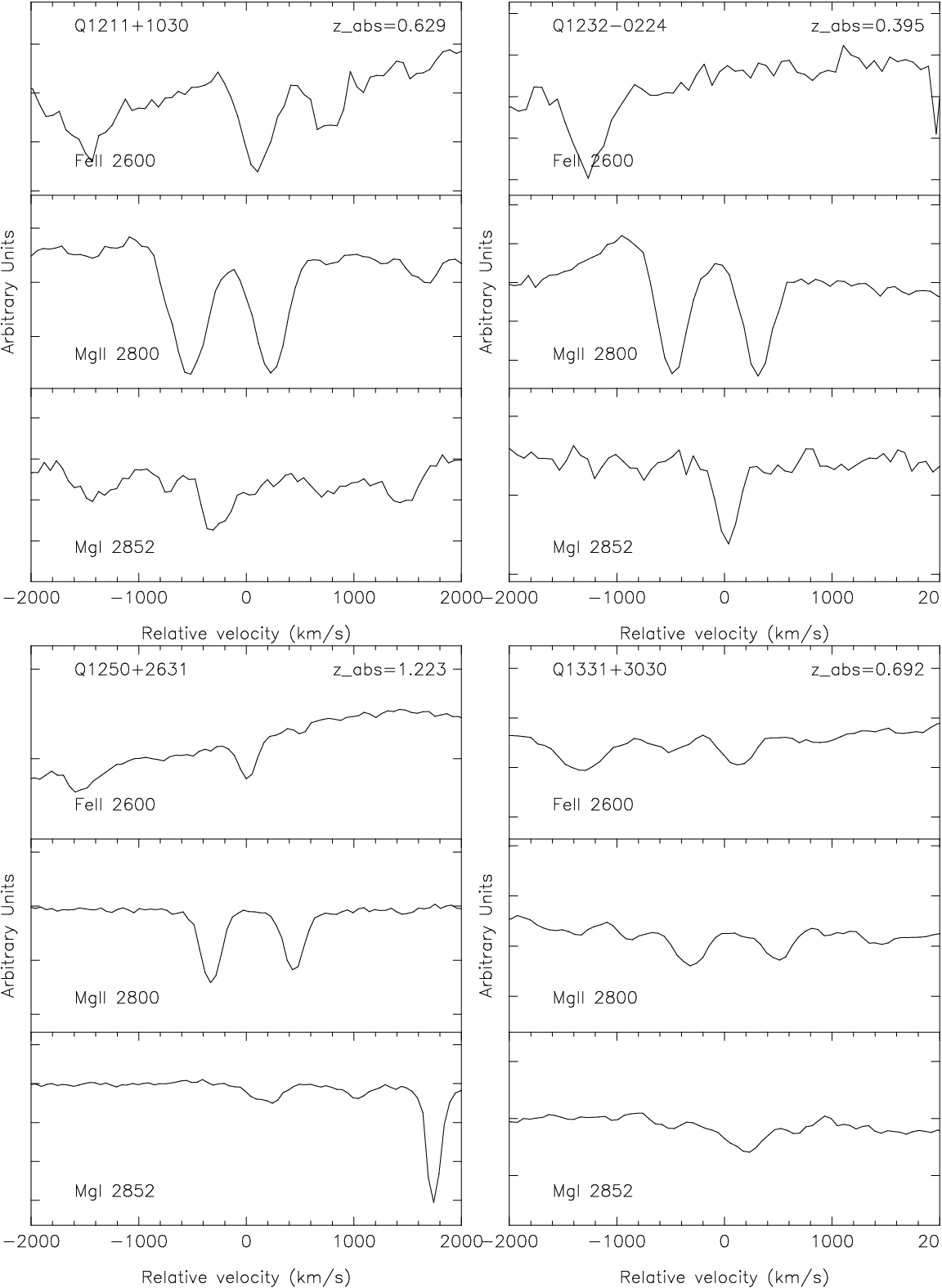}
\vspace{1.cm}		  
   \includegraphics[width=70mm]{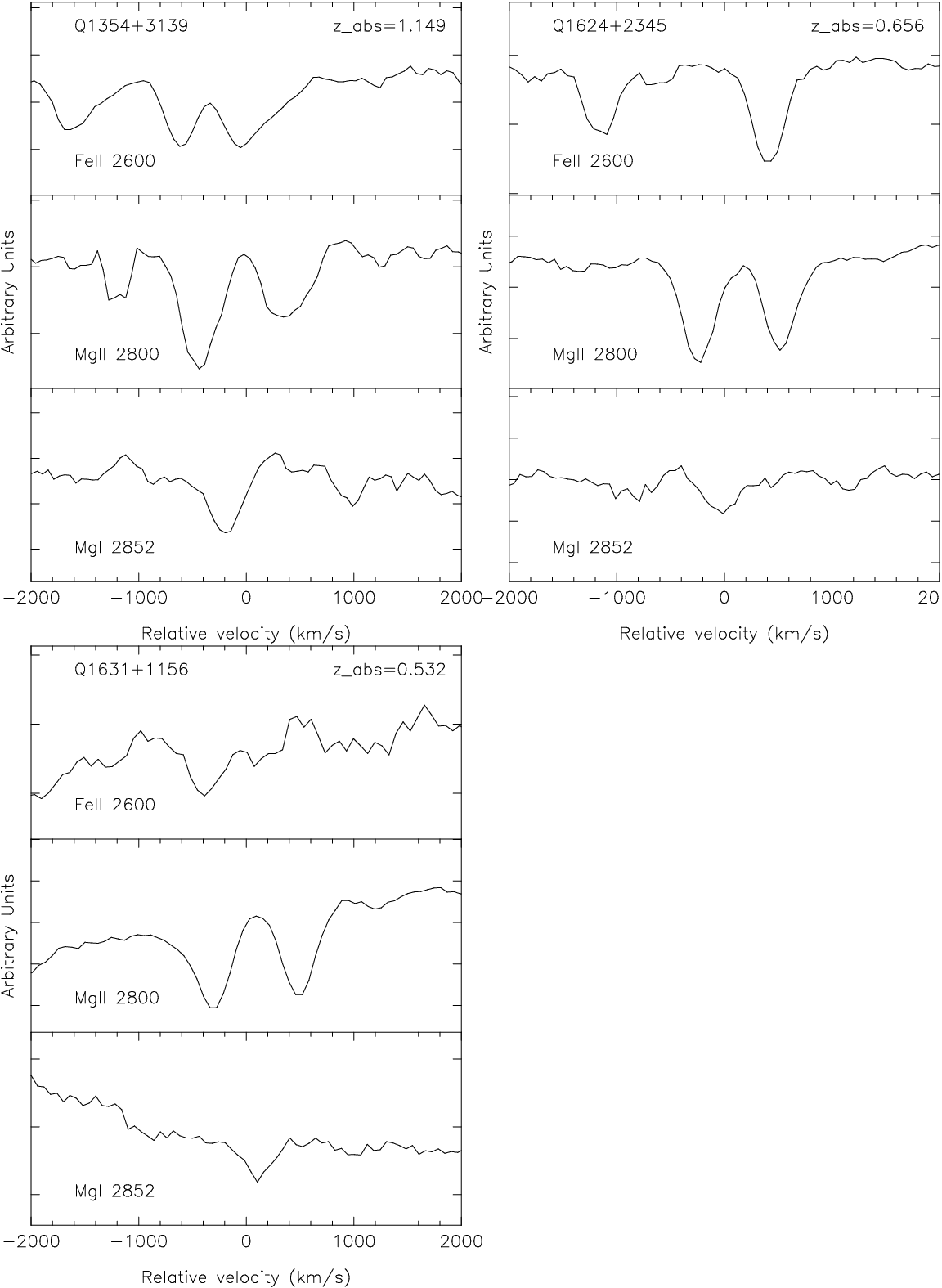}
\caption{\f, \M\ and \m\ measured as part of our survey. Thirteen of 
these systems were observed at the TNG while two additional
absorptions were measured in an archival HST spectrum
(Q0238$+$1636/0235$+$164) and in the commissioning ESO/UVES
spectrum, degraded to the TNG resolution (Q1106$-$1821/HE1104$-$1805).}
\label{f:MgII}
\end{figure*}

\section{The Data}

\subsection{\M\ Measurements}

In order to build the so-called ``classical'' sample, we have acquired
optical spectra covering the \f\ 2600, \M\ 2796, \M\ 2803 and \m\ 2852
features of known DLAs/sub-DLAs with well constrained \nhi\ column
densities. The observations were carried out in service mode with the
DOLORES spectrograph at the 3.58 m Telescopio Nazionale Galileo (TNG)
located at Roque de Los Muchachos Observatory in the Canary
Islands. High signal-to-noise optical spectrophotometry was obtained
covering the appropriate wavelength range so as to study the \M\ and
other associated metal lines in $z_{\rm abs}<1.7$ quasar absorbers,
the exact range depending on the grism used for the observations. A
journal of the observations is presented in 
Table 1.

Thirteen quasars were observed between February and April 2003. The
integrating times are all of a total of 1800 seconds regardless of the
brightness of the object. Only in one case (Q1354$+$3139/Q1351$+$318),
the weather conditions were poor and the resulting spectrum thus
suffers low signal-to-noise ratio. A combination of high and
low-resolution modes of the DOLORES spectrograph is used. The camera
is equipped with a Loral thinned and back-illuminated CCD with 2048
$\times$ 2048, 15 $\mu$ pixels. Grisms 1, 5, 6 and 7 (LR$-$B, HR$-$B,
HR$-$V and HR$-$R respectively) are used for the
observations. Therefore the dispersion of the spectra ranges from 0.8
to 2.8 \AA\space pixel$^{-1}$. All the observations were taken with a
slit width of 1.0 -- 1.5 arcsec.

The data reduction is done using the IRAF~\footnote{IRAF is
distributed by the National Optical Astronomy Observatories, which are
operated by the Association of Universities for Research in Astronomy,
Inc., under cooperative agreement with the National Science
Foundation.} software package. The bias frames are so similar over
successive nights that a master `zero' frame for each month is created
using the IMCOMBINE routine. After trimming the data, they are zero
corrected using CCDPROC. Similarly a single flat-field frame is
produced by taking the median of the flats. The overall background
variation across this frame is removed using IMSURFIT to produce an
image to correct for the pixel-to-pixel sensitivity variation of the
data. The task APALL is used to extract 1-D multi-spectra from the 2-D
frames of the quasars and spectrophotometric standard stars, while a
cut through the 2-D frames of the Arc served as an extraction
procedure. The APALL routine estimates the sky level by model fitting
over specified regions on either side of the spectrum. In the case
where the object is rather faint on the 2-D frame (i.e. for
Q0238$+$1636/0235$+$164), different exposures are IMCOMBINEd before
extraction with APALL. The spectra are then wavelength calibrated
using Ar, Ne or He arcs according to the grism used. We pay particular
attention to this step of the data processing, since it is crucial for
the later measurements of the metal equivalent widths.
Given that the objects have magnitudes ranging from 15 to 19 but
identical exposure time, the signal-to-noise ratios of the resulting
spectra vary from one object to another.

\begin{table*}
\begin{center}
\caption{Observing details for the spectra retrieved from public archives and used for \nhi\ measurements.}
\begin{tabular}{llcccllcc}
\hline
Coordinate & Alternative &$z_{\rm abs}$ &$\lambda_{\rm obs}$ &\logh\ &HST/ESO &Inst &Gratings &Selection \\ 
Name & Name & & & &prog ID (PI) & & &Method\\
\hline
Q0424$+$0204    &PKS0421$+$019       &1.638  &3207	&19.01$\pm$0.15 &6577 (Rao) 	&FOS	&270	&\M \\ 
Q0427$-$1302    &PKS0424$-$13        &1.408  &2927	&19.43$\pm$0.15 &6577 (Rao) 	&FOS	&270	&\M \\ 
Q0427$-$1302    &PKS0424$-$13        &1.562  &3114	&19.35$\pm$0.10 &6577 (Rao) 	&FOS	&270	&\M \\ 
Q0826$-$2230    &PKS0823$-$22        &0.910  &2322	&19.38$\pm$0.10 &6577 (Rao) 	&FOS	&270	&\M \\ 
Q0938$+$4129    &Q0935$+$417         &1.373  &2885	&20.45$\pm$0.10 &6237 (Beaver)   &FOS	&270	&\h \\ 
Q1106$-$1821$^a$&HE1104$-$1805       &1.662  &3236	&20.75$\pm$0.10 &60.A-9022        &UVES	&...	&\h \\ 
Q1325$+$6515    &4C65.15             &1.610  &3173	&19.76$\pm$0.10 &6577 (Rao) 	&FOS	&270	&\M \\ 
Q1331$+$4101    &PG1329$+$412        &1.282  &2774	&19.86$\pm$0.10 &6577 (Rao) 	&FOS	&270	&\M \\ 
Q1331$+$4101    &PG1329$+$412        &1.601  &3162	&19.33$\pm$0.15 &6577 (Rao) 	&FOS	&270	&\M \\ 
Q1429$+$4747    &PG1427$+$480        &0.221  &1484	&19.73$\pm$0.10 &6781 (Wills)    &FOS	&130	&\M \\ 
Q2131$-$1207    &Q2128$-$123/PHL1598 &0.430  &1738	&19.55$\pm$0.10 &4581 (Bahcall)  &FOS	&190	&\h \\ 
\hline
\end{tabular}

\vspace{0.1cm}

$^a$ UVES commissioning spectrum used for both \logh\ and \M\ measurements.\\
\end{center}
\label{t:add}
\end{table*}

\vspace{.5cm}
In addition to these new data, we thoroughly checked published
literature and public data sets for any other spectra of sufficient
quality for us to estimate equivalent widths of metal features. \f\
2600, \M\ 2796, \M\ 2803 and \m\ 2852 lines found in such a way are
only kept if they are of comparable quality to the one obtained with
our own TNG survey. Only two spectra met those criteria: the UVES
commissioning spectrum of Q1106$-$1821/HE1104$-$1805 and the HST/STIS
spectra of Q0238$+$1636/0235$+$164 taken as part of program 7294 (PI:
Cohen). In order to obtain a uniform set of data we degrade the UVES
spectrum of Q1106$-$1821/HE1104$-$1805 to the TNG data resolution
before undertaking the equivalent width measurements.

\vspace{.5cm}
The \f\ 2600, \M\ 2796, \M\ 2803 and \m\ 2852 absorption features of
all the 15 absorbers are shown in Figure~\ref{f:MgII}. We compute the
equivalent widths of the metal lines for the 15 spectra at our
disposal. One object in our TNG sample (Q1001$+$5553/Q0957$+$561) does
not cover the \f\ 2600 line while on several occasions \f\ 2600 and/or
\m\ 2852 are not detected at the redshift expected for the
absorber. It can be seen from Figure 1 that the \M\ doublet is
resolved in all the spectra used for this study.

\subsection{\nhi\ Measurements}

In addition to the systems presented in the section above, we search
the literature for suitable \logh\ measurements. In order to gather a
significant sample, we look for any absorber down to the sub-DLA
definition (\logh$>$19.0) and with $z_{\rm abs}<1.7$. We only include
\nhi\ measurements derived from high signal-to-noise HST (STIS or FOS)
spectra. We therefore exclude measurements based on IUE spectra or
\nhi\ estimates from 21cm observations or other even more indirect
methods. Furthermore, we thoroughly check ESO and HST archives for any
available spectra for which the relevant measurements were not
available in the literature. Eleven \nhi\ measurements are possible
thanks to HST/FOS spectra recovered from the HST/ESO archives. The
UVES commissioning spectrum of Q1106$-$1821/HE1104$-$1805
provides a further measurement of \nhi. In particular, we extend the
Rao \& Turnshek sample by including 7 sub-DLAs which were
\M-selected but that these authors rejected because they had \logh$<$20.3. We also include the two \M-selected absorbers mentioned by 
Churchill (2001) and for which the \nhi\ measurements could be made
(Q1429$+$4747/PG1427$+$480 and Q0441$-$4313/PKS0439$-$433).

In order to measure \nhi\ column densities of the 11 absorbers
available to use, we normalise the spectra in the region of interest
using spline functions. We then fit Voigt profiles using the FITLYMAN
package within MIDAS (Fontana \& Ballester 1995). We note that for
several HST/FOS spectra, the absorbing feature does not reach the zero
flux level in its center, even if the line is clearly saturated. We
use a convolution of the Voigt profile together with the instrumental
line spread function to estimate whether this effect is real or
not. Only in a few cases do we find it to be an artifact and we
correct the spectrum by subtracting a few percent from the continuum
level as is standard practice (Boiss\'e
\e\ 1998; Rao \& Turnshek 2000). The typical resulting error bar in \logh\
measurements is 0.10 and never exceeds 0.15, but does not include error
in the continuum placement. The resulting fits are shown in
Figure~\ref{f:HI} for the eleven systems, while information on the
observing details is provided in Table 2.

\begin{figure*}
   \includegraphics[height=210mm]{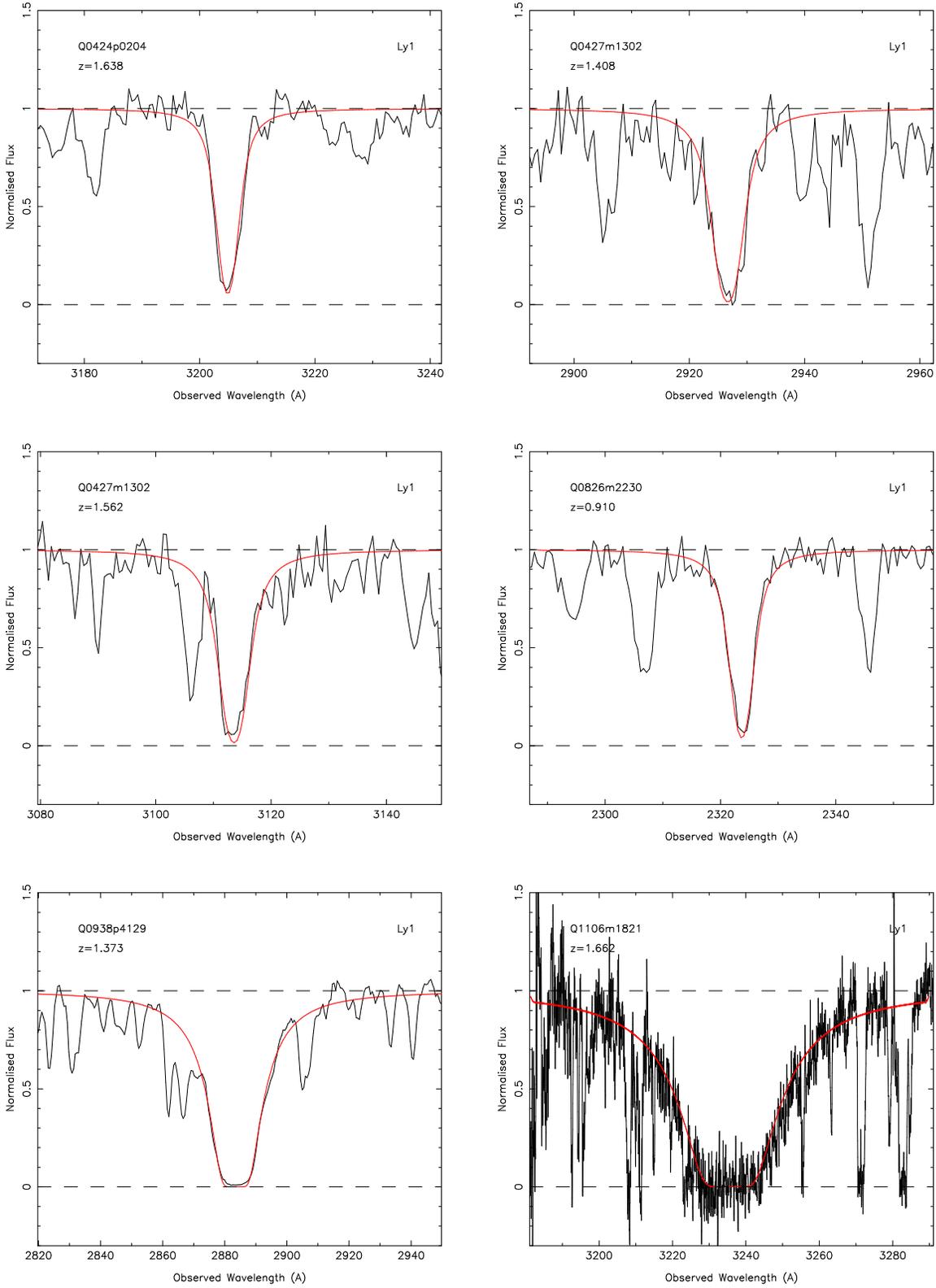}
\caption{Voigt profile fits of \nhi\ retrieved from public archives. Ten 
of these spectra are recovered from the HST/ESO archive while the
remaining system is measured in the commissioning ESO/UVES
spectrum (Q1106$-$1821/HE1104$-$1805).}
\setcounter{figure}{1}
\label{f:HI}
\end{figure*}

\begin{figure*}
   \includegraphics[width=160mm]{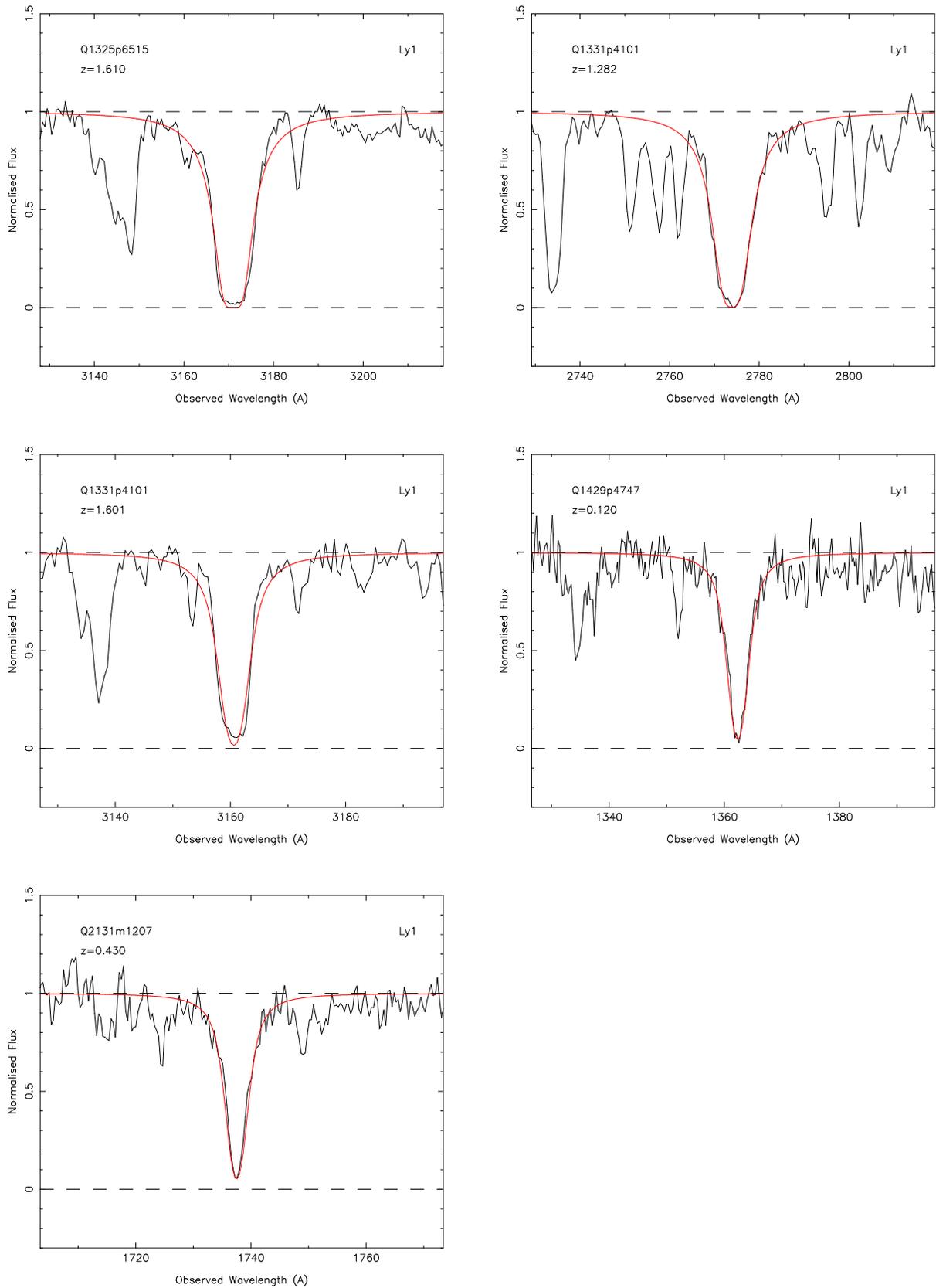}
\caption{{\it continued}}
\label{f:HI}
\end{figure*}

\section{\M\ Equivalent Widths Distributions}

In this section, we study the distribution of the equivalent widths of
metals in \M-selected absorbers as compared to that of more classical
DLAs/sub-DLAs. The latter sample is further divided into two
sub-samples: on one hand absorbers selected by their \h\ feature and
on the other hand systems selected at 21cm or in one case, because a
galaxy is known and later turns out to be a DLA.

\subsection{Samples Definition}

The method developed by Rao \& Turnshek (2000) makes use of the
empirical fact that all DLAs studied are found to have a corresponding
\M\ metal-like absorption (Rao 1994; Rao, Turnshek \& Briggs 1995).
In their study, the authors select a sample of 243 $z_{\rm abs}<1.65$, EW
\M\ (2796)$>0.3$\AA\ \M\ absorbers taken from various sources in the
literature. A fraction of these absorbers (87) already has a spectrum
as part of the HST archives (48 absorbers) while others are observed
as part of their own HST observing program (39 \M\ doublets). Using
these data, they measure whether a DLA is present or not at the
expected absorbing redshift.

In the present study, the emphasis is on constructing well defined
DLA/sub-DLA samples: one where the absorbers are \M-selected, and
another for the remaining systems, discovered predominantly, but not
only, on their \h\ feature. Therefore the latter also includes
absorbers selected at 21 cm or a known galaxy which is further found
to produce a DLA signature in a quasar spectrum. Moreover, we extend
the work of Rao \& Turnshek (2000) (as well as our literature search
for ``classical'' systems) down to the sub-DLA definition in order to
get a more statistically significant sample of absorbers. This is done
by measuring the \h\ column densities of Rao \& Turnshek's (2000) HST
program for which \M\ absorbers were detected. 

The resulting {\bf \M-selected sample} in the present study is thus
larger than the original work from Rao \& Turnshek (2000). It is
composed of 21 systems defined as follows:

\begin{itemize}

\item {\bf 7} DLAs newly discovered in Rao \& Turnshek (2000).

\item {\bf 7} sub-DLAs in the same observing program and fitted by us.

\item {\bf 1} of the 4 very low redshift systems of Churchill (2001) 
for which \logh\ is reported in the literature
(Q0441$-$4313/PKS0439$-$433) and {\bf 1} other for which \logh\
is determined by us (Q1429$+$4747/PG1427$+$480).

\item {\bf 1} sub-DLA (Q0100$+$0211/Q0058$+$019PHL938) selected by 
Rao \& Turnshek (2000) and {\bf 1} other DLA found by the same group
with their new HST program (Q1631$+$1156/Q1629$+$120).

\item {\bf 2} systems (Q0456$+$0400/Q0454$+$039 and Q1211$+$1030/Q1209$+$107) 
observed during the HST program 5351 (PI: Bergeron) based on their
metal features in optical spectra (see Boiss\'e \e\ 1998).

\item {\bf 1} DLA (Q1624$+$2345,Q1622$+$238/3C336) selected thanks to its \M\ 
signature and observed during HST program 5304 (PI: Steidel).

\end{itemize}

It should be noted that most of the quasar spectra with DLAs in the
HST archives are clearly taken because the observer expects the
absorber to be present. As an example, for the system along
(Q0304$-$2211/EX$-$PKS0302$-$223), the HST proposal abstract (ID:
6224; PI: Burbidge) states that ``This data set will yield high
quality absorption line data for studies of the damped Lyman alpha
absorption system''. Therefore, such absorbers are initially not
selected on the basis of their \M\ feature and we do not include them
in our
\M-selected sample even if they have been ``re-discovered'' in the
archive search of Rao \& Turnshek (2000). Rather, we define the
\M-selected sample as all systems which have been newly discovered
based on their metal signature in optical quasar spectra. 

\vspace{1cm}

The remaining systems form the {\bf classical sample} which can be
defined as follows:

\begin{itemize}

\item {\bf 10} DLAs/sub-DLAs directly discovered thanks to their \h\ 
signature. These systems also form the so-called {\it \h-selected}
sample (see Section 4).

\item {\bf 4} absorbers selected at 21cm and with \nhi\ measured in UV 
spectra (Q0238$+$1636/Q0235$+$164; Q0813$+$4813/Q0809$+$483/3C196;
Q1232$-$0224/PKS1229$-$021; Q1331$+$3030/Q1328$+$307/3C286).

\item {\bf 1} system for which the quasar-galaxy association is 
known prior to the HST observations (Q1215$+$3309/TON1480).

\end{itemize}

Altogether, the number of systems studied here is 36. A summary of
the resulting sample is shown in Table~\ref{t:summary} together with
the properties of both the quasar itself and its absorbers. It should
be emphasized once more that we only include systems for which {\it
both} \nhi\ and metal equivalent widths can be securely
recovered. Hence, while during the course of our literature search we
have known of as many as 93 DLAs/sub-DLAs with $z_{\rm abs}<1.7$, our
criterion reduces the current sample to 36 absorbers, 21 of which are
\M-selected. In particular, we choose not to include the two \M-selected 
DLAs (\logh=21.16 and \logh=21.41) recently reported by Turnshek \e\
(2004) since these initial results might not be representative of the
most recent HST UV spectroscopic survey undertaken by their group.


\begin{landscape}
\begin{table}
\caption{Summary of the properties of the 36 $z_{\rm abs}<1.7$ absorbers (21 of which are \M-selected) making up our complete sample.}
\begin{tabular}{llccccccccccc}
\hline \hline
Coordinate &Alternative &$z_{\rm em}$ &Mag &$z_{\rm abs}$ &\logh\ &Ref &$EW_{\rm rest}$ &$EW_{\rm rest}$ &$EW_{\rm rest}$
&$EW_{\rm rest}$ &Ref &Selection \\ Name &Name && & &[atom cm$^{-2}$]
& \logh\ &\f 2600 &\M 2796 &\M 2803 &\m 2852 &$EW_{\rm metals}$
&Method \\
\hline			                       	       	      	       	     	   	       
Q0100$+$0211  &Q0058$+$019/PHL938  &1.954  &17.2  &0.613  &20.08 &(3)  &1.39    &1.70    &1.51    &0.35  &(5)  &\M \\ 
Q0217$+$0144  &Q0215$+$015$^a$     &1.715  &18.3  &1.345  &19.57 &(4)  &2.06    &2.57    &1.63    &0.37  &(1)  &\h \\ 
Q0238$+$1636  &Q0235$+$164$^a$     &0.940  &19.0  &0.524  &21.70 &(3)  &2.13    &2.42    &3.28    &0.79  &(1)  &21 cm\\ 
Q0251$+$4315  &Q0248$+$430         &1.310  &17.4  &0.394  &21.56 &(5)  &1.03    &1.86    &1.42    &0.70  &(5)  &\M \\ 
Q0304$-$2211  &EX/PKS0302$-$223    &1.400  &16.0  &1.009  &20.36 &(12) &0.63    &1.16    &0.96    &0.18  &(5)  &\h \\ 
Q0424$+$0204  &PKS0421$+$019       &2.055  &17.0  &1.638  &19.01 &(1)  &$<$0.4  &0.35    &0.28    &$<$0.2&(5)  &\M \\ 
Q0427$-$1302  &PKS0424$-$13        &2.166  &17.5  &1.408  &19.43 &(1)  &0.44    &0.55    &0.35    &$<$0.3&(5)  &\M \\ 
Q0427$-$1302  &PKS0424$-$13        &2.166  &17.5  &1.562  &19.35 &(1)  &$<$0.2  &0.38    &0.39    &$<$0.2&(5)  &\M \\ 
Q0441$-$4313  &PKS0439$-$433       &0.593  &16.4  &0.101  &20.00 &(3)  &0.91    &2.32    &1.99    &0.54  &(6)  &\M \\
Q0456$+$0400  &Q0454$+$039         &1.345  &16.5  &0.859  &20.69 &(7)  &1.21    &1.48    &1.49    &0.30  &(1)  &\M \\ 
Q0741$+$3112  &OI363/Q0738$+$313   &0.635  &16.1  &0.221  &20.90 &(3)  &$<$0.6  &0.52    &0.50    &$<$0.2&(5)  &\M \\ 
Q0813$+$4813  &Q0809$+$483/3C196   &0.871  &17.8  &0.437  &20.80 &(8)  &1.93    &1.97    &2.06    &1.38  &(1)  &21 cm \\ 
Q0826$-$2230  &PKS0823$-$22        &0.920  &16.2  &0.910  &19.38 &(1)  &...     &1.15    &0.68    &$<$0.4&(5)  &\M \\ 
Q0830$+$2410  &Q0827$+$243         &0.939  &17.3  &0.518  &20.30 &(3)  &1.90    &2.90    &2.20    &...   &(5)  &\M \\ 
Q0853$+$4349  &Q0850$+$440         &0.514  &16.4  &0.164  &19.81 &(9)  &$<$0.2  &$<$0.2  &0.43    &...   &(9)  &\h \\ 
Q0937$+$7301  &Q0933$+$733         &2.528  &17.0  &1.478  &21.62 &(5)  &0.76    &0.95    &1.15    &$<$0.3&(5)  &\M \\ 
Q0938$+$4129  &Q0935$+$417         &1.980  &16.2  &1.373  &20.45 &(1)  &0.61    &1.04    &1.02    &...   &(1)  &\h \\ 
Q0954$+$1743  &Q0952$+$179         &1.472  &17.2  &0.239  &21.32 &(5)  &...     &0.63    &0.79    &$<$0.4&(5)  &\M \\ 
Q1001$+$5553  &Q0957$+$561$^b$     &1.414  &16.7  &1.391  &20.28 &(10) &...     &2.25    &1.93    &0.38  &(1)  &\h \\ 
Q1106$-$1821  &HE1104$-$1805$^b$   &2.305  &16.7  &1.662  &20.75 &(1)  &0.73    &0.95    &0.90    &0.32  &(1)  &\h \\ 
Q1124$-$1705  &HE1122$-$1649       &2.400  &16.5  &0.680  &20.45 &(3)  &1.23    &1.72    &1.57    &0.10  &(1)  &\h \\ 
Q1130$-$1449  &PKS1127$-$14        &1.184  &16.9  &0.313  &21.71 &(3)  &1.14    &2.21    &1.90    &1.14  &(5)  &\M \\ 
Q1211$+$1030  &Q1209$+$107         &2.187  &17.6  &0.629  &20.20 &(7)  &1.11    &2.54    &2.35    &0.56  &(1)  &\M \\ 
Q1215$+$3309  &TON1480             &0.614  &17.0  &0.004  &20.34 &(11) &0.68    &1.21    &0.86    &0.20  &(11) &Gal \\ 
Q1232$-$0224  &PKS1229$-$021       &1.045  &17.6  &0.395  &20.75 &(7)  &1.73    &2.00    &1.82    &0.64  &(1)  &21 cm \\ 
Q1250$+$2631  &Q1247$+$267         &2.038  &15.8  &1.223  &19.87 &(2)  &0.23    &0.51    &0.43    &0.24  &(1)  &\h \\ 
Q1325$+$6515  &4C65.15             &1.624  &17.5  &1.610  &19.76 &(1)  &0.88    &2.20    &1.85    &0.16  &(5)  &\M \\ 
Q1331$+$3030  &Q1328$+$307/3C286   &0.849  &17.2  &0.692  &21.19 &(7)  &0.21    &0.31    &0.25    &0.36  &(1)  &21 cm \\ 
Q1331$+$4101  &PG1329$+$412        &1.937  &17.2  &1.282  &19.86 &(1)  &...     &0.49    &0.31    &$<$0.3&(5)  &\M \\ 
Q1331$+$4101  &PG1329$+$412        &1.937  &17.2  &1.601  &19.33 &(1)  &$<$0.2  &0.70    &0.35    &$<$0.2&(5)  &\M \\ 
Q1354$+$3139  &Q1351$+$318         &1.326  &17.4  &1.149  &20.23 &(2)  &1.12    &2.10    &1.53    &0.75  &(1)  &\h \\ 
Q1357$+$2537  &Q1354$+$258         &2.032  &18.5  &1.420  &21.54 &(2)  &0.55    &0.61    &0.50    &0.20  &(5)  &\M \\ 
Q1429$+$4747  &PG1427$+$480        &0.221  &16.3  &0.120  &19.73 &(1)  &0.40    &0.78    &0.66    &$<$0.5&(6)  &\M \\ 
Q1624$+$2345  &Q1622$+$238/3C336   &0.927  &17.5  &0.656  &20.36 &(3)  &1.13    &1.52    &1.27    &0.39  &(1)  &\M \\ 
Q1631$+$1156  &Q1629$+$120         &1.795  &18.5  &0.532  &20.70 &(3)  &0.68    &1.55    &1.34    &0.43  &(1)  &\M \\ 
Q2131$-$1207  &Q2128$-$123/PHL1598 &0.501  &15.5  &0.430  &19.55 &(1)  &0.27    &0.41    &0.37    &0.10  &(5)  &\h \\ 
\hline\hline		       
\end{tabular}

\vspace{0.1cm}

$^a$ Quasar with Broad Absorption Lines (BAL).\\
$^b$ Gravitationally lensed quasar.\\
\vspace{0.1cm}
{\bf References:} \\
(1)~This work; 
(2)~Pettini \e\ 1999;
(3)~Chen \& Lanzetta 2003;
(4)~Ledoux \e\ 2002;
(5)~Rao \& Turnshek 2000;
(6)~Churchill 2001;
(7)~Boiss\'e \e\ 1998;
(8)~Boiss\'e \e\ 1990;
(9)~Lanzetta 1997;
(10)~Zuo \e\ 1997;
(11)~Miller \e\ 1999;
(12)~Turnshek \& Rao 2002;
\label{t:summary}
\end{table}
\end{landscape}

\newpage

\begin{figure}
   \includegraphics[width=80mm,angle=-90]{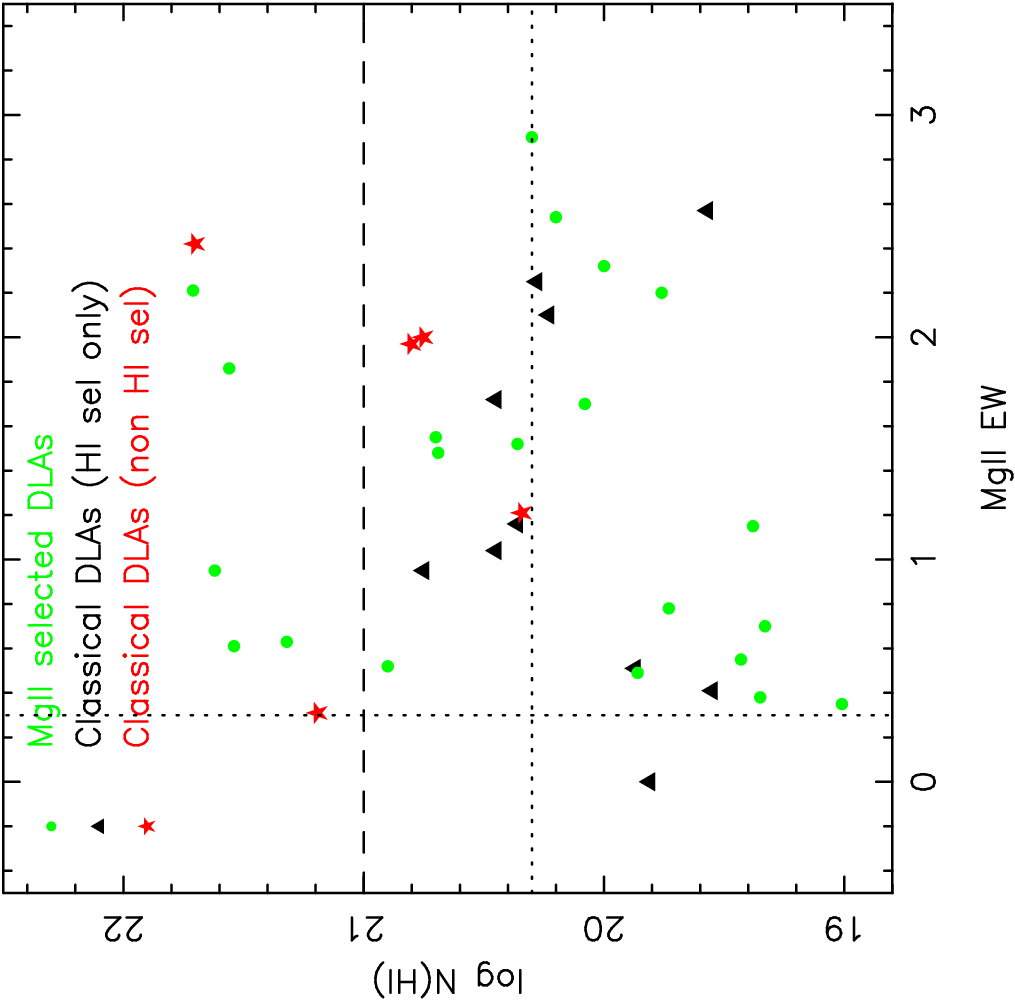}
\vspace{1.cm}		  
   \includegraphics[width=80mm,angle=-90]{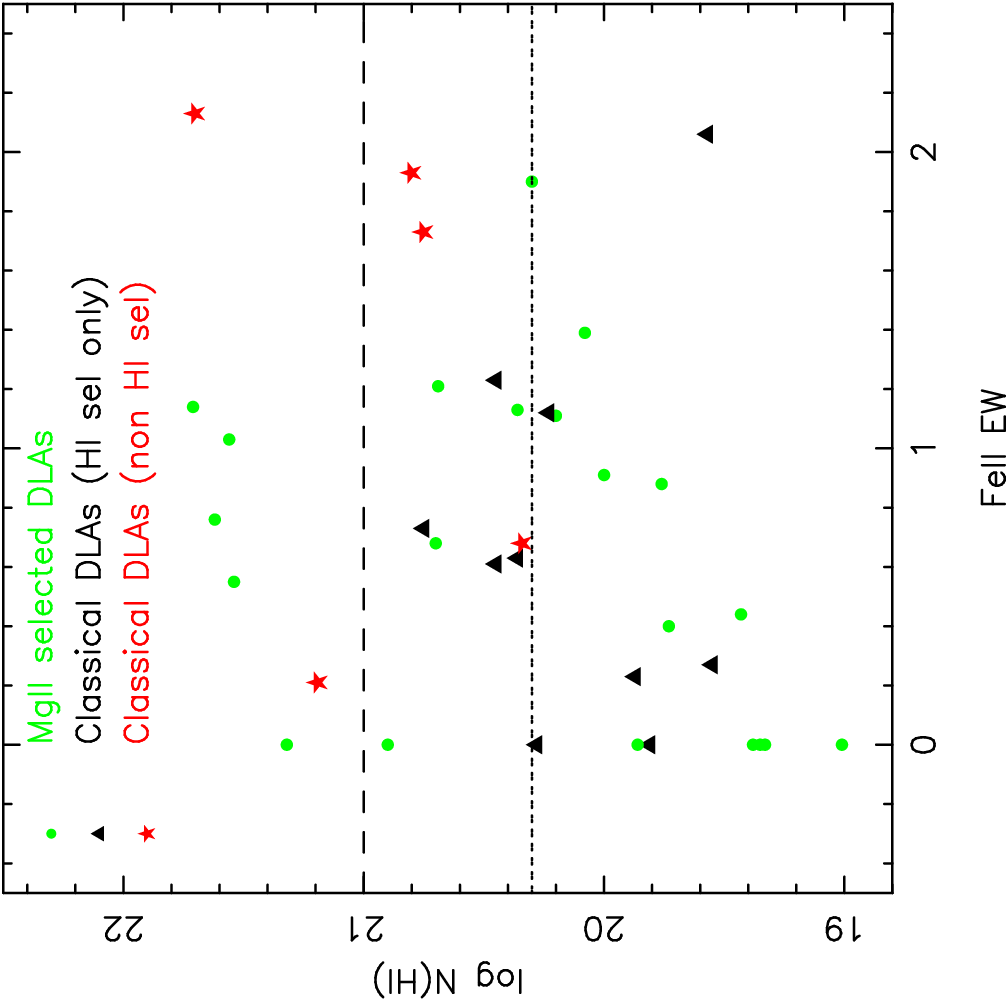}
\caption{{\it Top panel:} \logh\ as a function of the  \M\ 2796 \AA\ 
rest equivalent widths. The dashed horizontal line corresponds to
\logh=21.0, while the dotted horizontal line is \logh=20.3. The dotted
vertical line is the \M\ rest equivalent width lower limit that Rao \&
Turnshek (2000) use as a criterion for their DLA search.{\it Bottom
panel:} Same for \f\ 2600. }
\label{f:RTSel}
\end{figure}

\begin{figure}
   \includegraphics[width=80mm,angle=-90]{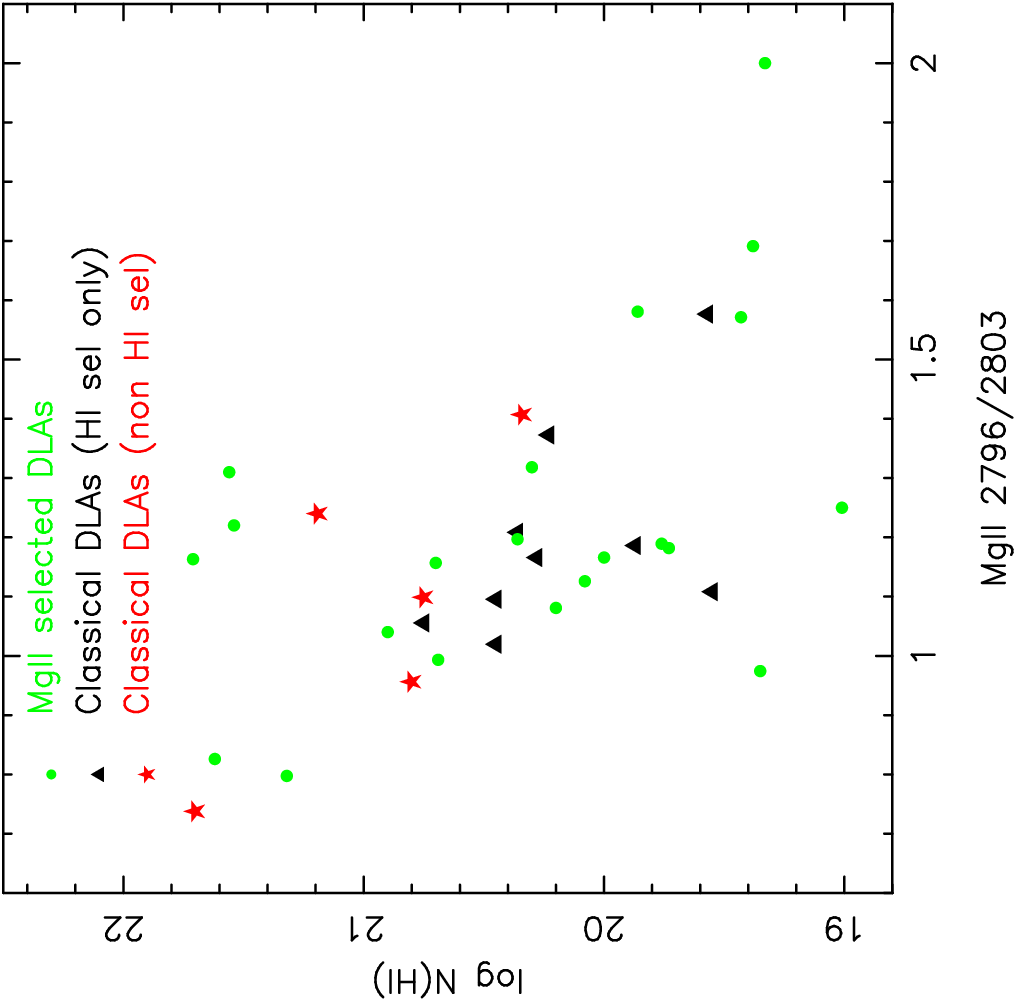}
\vspace{1.cm}		  
   \includegraphics[width=80mm,angle=-90]{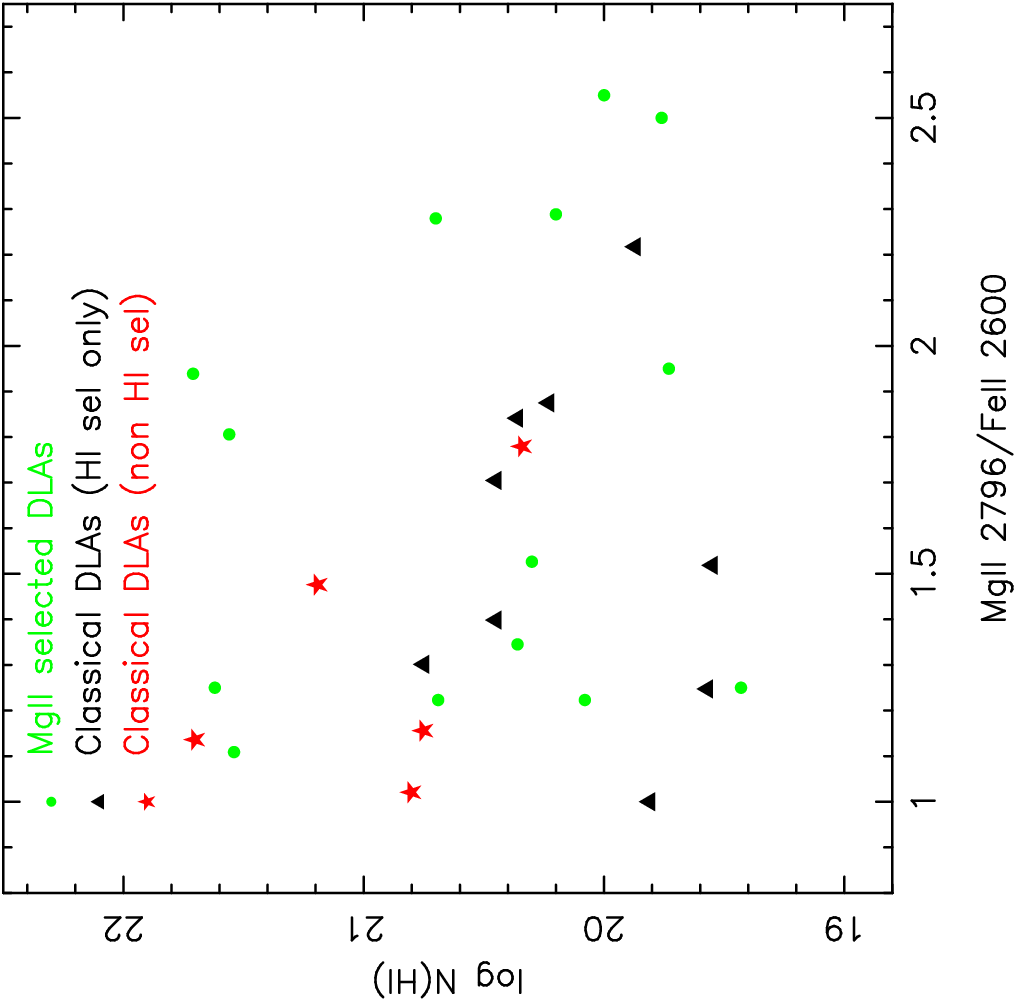}
\caption{{\it Top panel:} \logh\ as a function of the ratio 
$EW_{\M(2796)}/EW_{\M(2803)}$. {\it Bottom panel:} \logh\ as a
function of the ratio $EW_{\M(2796)}/EW_{\f(2600)}$.} 
\label{f:ratio}
\end{figure}

\subsection{Metal Equivalent Widths}

We compare the \M/\f\ EW distribution of \M-selected and classical
systems. A simple picture would have Mg~{\sc ii} equivalent widths
correlating with the column densities. Indeed, Churchill \e\ (2000)
observe that at $z<1.6$ the \M\ equivalent widths are dominated by
kinematic spreads. They propose that these could be directly
correlated to column density, implying in turn a correlation between
EW and \nhi.

In order to address this issue we plot in Figure~\ref{f:RTSel} \M\
2796 (top panel) and \f\ 2600 (bottom panel) equivalent widths for
each of the 36 systems. Upper limits and non-detections are plotted at
zero. 

In their work, Rao \& Turnshek (2000) look for DLAs only when
\M\ 2796 has a rest equivalent width EW$>$0.3\AA. This is illustrated
in the top panel of Figure~\ref{f:RTSel} by the vertical dotted
line. It can be seen that indeed such a criterion is appropriate for
selecting DLAs with \logh$>20.3$ cm$^{-2}$ (which is precisely
the aim of these authors), while any higher \M\ EW threshold will miss
a fraction of the absorbers. This DLA definition is represented by the
horizontal dotted line. In the present work, we extended both
classical and \M-selected samples to the sub-DLA definition in order
to get larger samples. We note that even down to lower column
densities, the rest equivalent width threshold set by Rao \& Turnshek
(2000) is still a good tracer of quasar absorbers.

Nevertheless, it is puzzling to notice that the majority of absorbers
with \logh$>21.0$ (horizontal dashed line) are
\M-selected, while fewer of these systems have $20.3 < \logh\ <
21.0$. Conversely, in general classical absorbers do not have such
high column densities and this regardless of the way they have been
selected (direct \h\ selection, 21 cm or known galaxy).

In Figure~\ref{f:ratio}, we plot the \logh\ column density as a
function of metal equivalent widths ratios. The top panel shows
$EW_{\M(2796)}/EW_{\M(2803)}$ while the bottom panel displays
$EW_{\M(2796)}/EW_{\f(2600)}$. We find again that both classical and
\M-selected absorbers span a similar parameter range for both these
quantities. 

\newpage

\section{\nhi\ Distributions}

\subsection{Comparison of the Two Samples}

\begin{figure}
   \includegraphics[width=60mm,angle=-90]{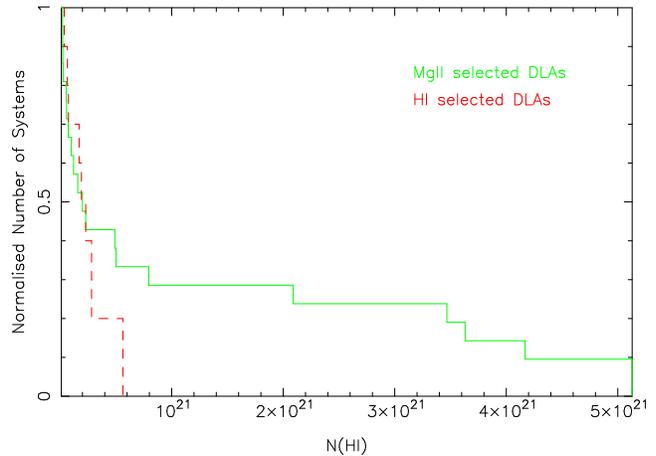}
\caption{Normalised cumulative distribution of \nhi\ for \h-selected and
\M-selected absorbers.
}
\label{f:cumu}
\end{figure}

In this section, we analyse the normalised cumulative \nhi\
distributions of the \M-selected sample and compare it with a purely
\h-selected sample. This distribution directly impacts
measurements of the cosmological mass density, $\Omega_{\rm HI}$.
Here, the \M-selected sample presented in section 3.1 is compared with
the sample of DLAs/sub-DLAs selected {\it only} on their \h. The
latter does not include absorbers selected at 21cm or otherwise but
solely \h-selected systems. Therefore the total sample is now composed
of 31 absorbers, 21 of which are \M-selected. Figure~\ref{f:cumu}
presents the normalised cumulative \nhi\ distributions for these. The
\M-selected curve is different from the other at the
high \nhi\ column density end. In particular, there are systems with
\logh$>$21.0 in the \M-selected sample while \h-selected systems are
never found to have \logh\ $>$ 20.8. This cannot be due to a bias in
the \h-selected sample, since that selection method should in
principle favour high \nhi\ column densities. A Kolmogorov-Smirnov
(KS) test shows that the two distributions are consistent at a 0.56
level of probability, but clearly we are dealing with small number
statistics.  In any case, the sheer fraction of high column density
systems will in turn affect the determination of $\Omega_{\rm HI}$
computed directly from the integration of the \nhi\ distribution
function.

Further, we find that by comparing systems with \logh$>$20.3 only,
none of the \h-selected DLAs have \logh$>$ 21 (out of 4), while 50\%
(5 out of 10) of the \M-selected DLAs have
\logh$>$ 21 (see Figure~\ref{f:RTSel}). Assuming a poisson
distribution, the probability of obtaining zero \h-selected DLAs at
\logh$>$ 21 when expected 2 (i.e. half of 4) is 13.5\%. Therefore
the absence of \h-selected \logh$>$ 21 systems might also be due
to statistical fluctuation. When all classical DLAs are taken into
account: only 25\% (2 out of 8) have \nhi\ $>$ 21 even though these
have been selected using different techniques. Already Rao \& Turnshek
(2000) note that the comparison of a sample of local gas-rich galaxies
studied in 21 cm emission ($<z>$=0) with \M-selected low redshift DLAs
($<z>$=0.5) show an unexpectedly high rate of occurrence of very large
column densities. This could be due to redshift evolution or the
selection methods or a combination of both.

Therefore, the \M-selected sample has a higher fraction of \logh$>$ 21 systems, but only marginally. Given the impact of this on
$\Omega_{\rm HI}$, we try to identify what could be the origin of such
a discrepancy by testing various scenarii in the following sections.

\begin{figure}
   \includegraphics[width=80mm,angle=-90]{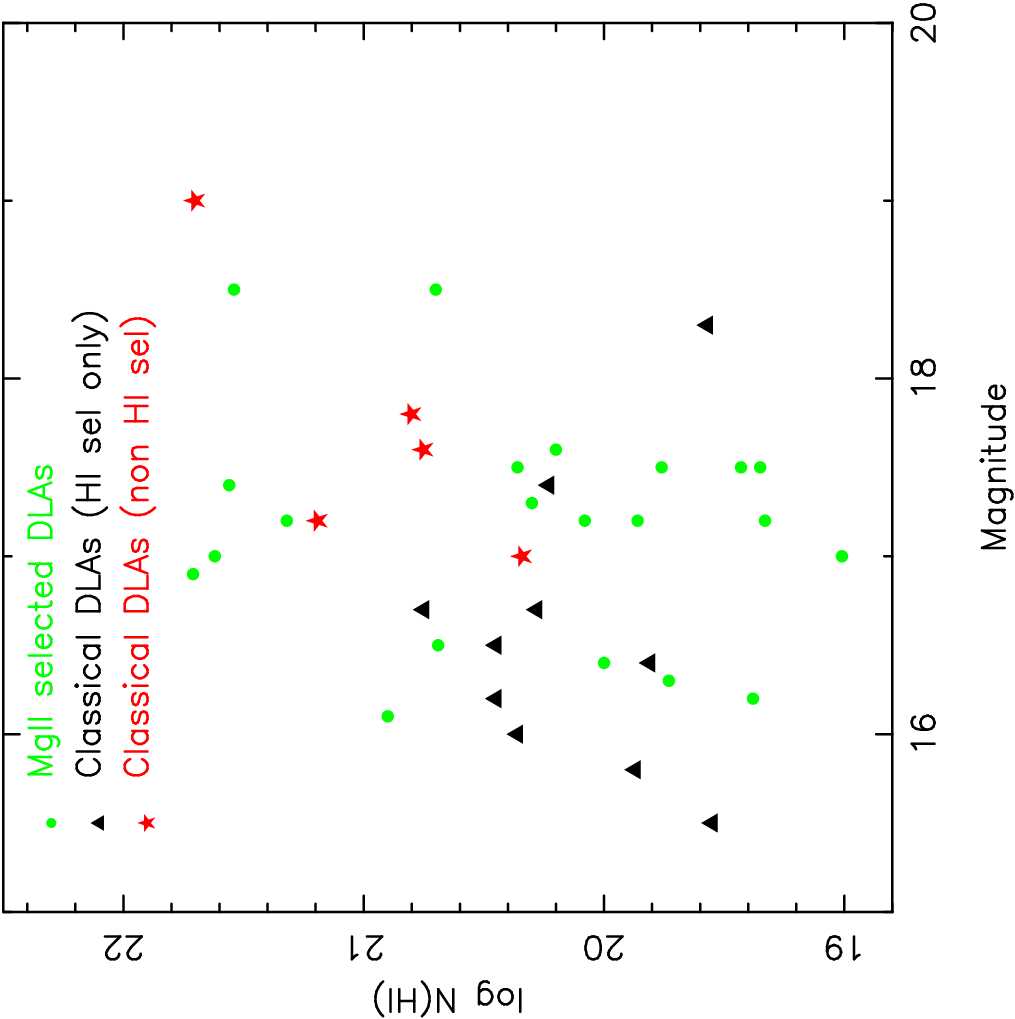}
\vspace{1.cm}		  
   \includegraphics[width=80mm,angle=-90]{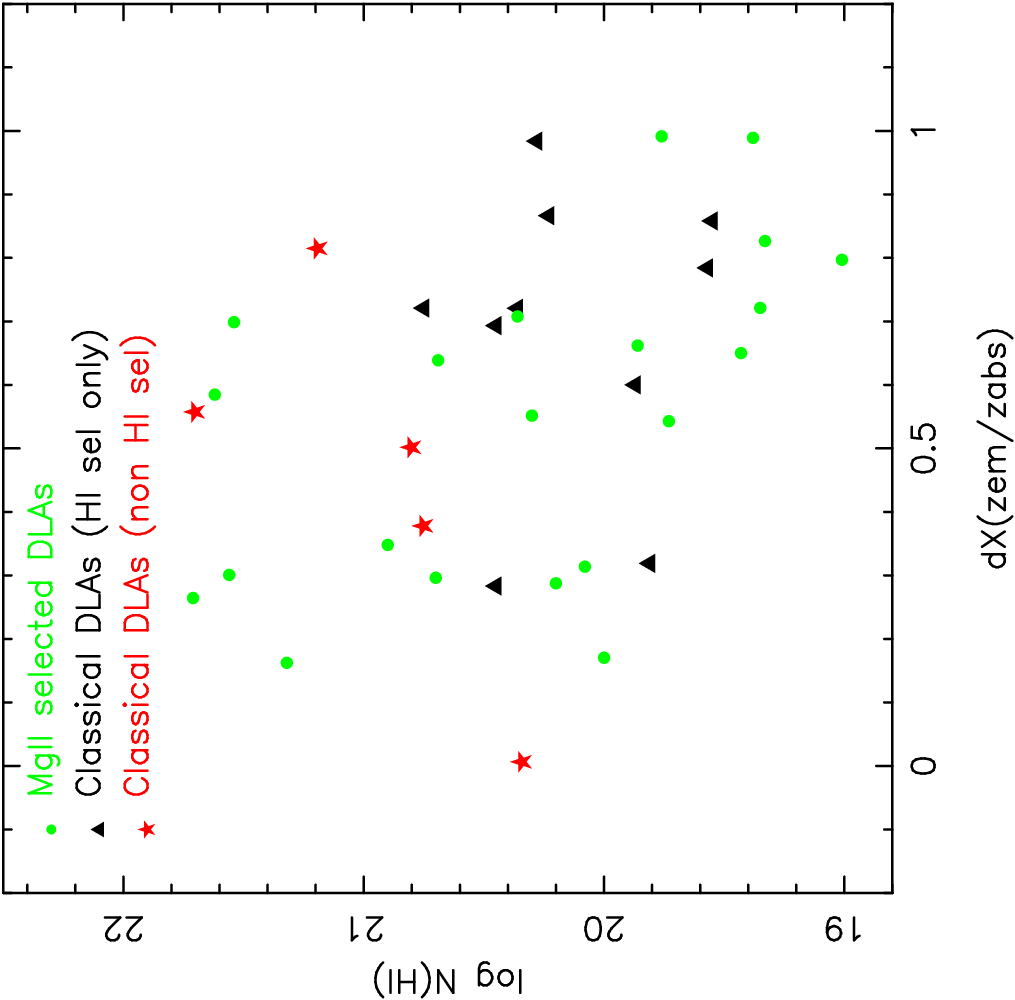}
\caption{{\it Top panel:} \logh\ as a function of the apparent background 
quasar magnitude. {\it Bottom panel:} \logh\ as a function of the
ratio of the physical distance to the absorber over the physical
distance to the quasar. According to gravitational lensing theory, the
quasar magnification is maximised when this ratio is 0.5.}
\label{f:GL}
\end{figure}

\subsection{Gravitational Lensing}

Another effect possibly at play could be gravitational lensing of the
background quasar by the absorber itself. This could amplify the
luminosity of the quasars and therefore make them preferentially
selected as they would be easier targets to observe. In order to check
whether gravitational lensing affects differently the two samples of
absorbers, we plot the apparent magnitude of quasars for the samples
of classical and \M-selected DLAs/sub-DLAs. Looking at the top panel
of Figure~\ref{f:GL}, we find no evidence that the two populations are
composed of quasars with different magnitude ranges. There might be
hints of a trend where higher column density absorbers are found in
fainter quasars, probably a signature of extinction by dust (Boiss\'e
\e\ 1998), but this result is not currently statistically significant.

In order to test further the strong lensing hypothesis, we perform
another test. It is known from gravitational lensing theory that the
magnification effects are maximised when the lensing object is midway
between the source and the observer {\it in physical space}. In our
given cosmology, the physical distances can be expressed as:

\begin{equation}
X(z) = \int_{0}^{z} (1 + z)^2 \left[(1 + z)^2 (1 +
0.3 z) - 0.7 z (2 + z) \right]^{-1/2}dz
\end{equation}

In the bottom panel of Figure~\ref{f:GL}, we plot the ratio of the
distance to the absorber over the distance to the quasar. The
configuration which would most favour the occurence of gravitational
lensing would have this ratio equal to 0.5. In our study, the two
samples of absorbers do not show any significant
difference. Therefore, gravitational lensing does not appear to be a
viable explanation for the observed discrepancy in \logh\ distributions
of the classical and \M-selected DLAs/sub-DLAs. This finding is
further supported by the results of Le Brun \e\ (2000) who have used
the DLA sample of Le Brun \e\ (1997) to show that the amplification
due to the presence of an absorber along the quasar line of sight does
not exceed 0.3 mag.

\begin{figure}
   \includegraphics[width=80mm,angle=-90]{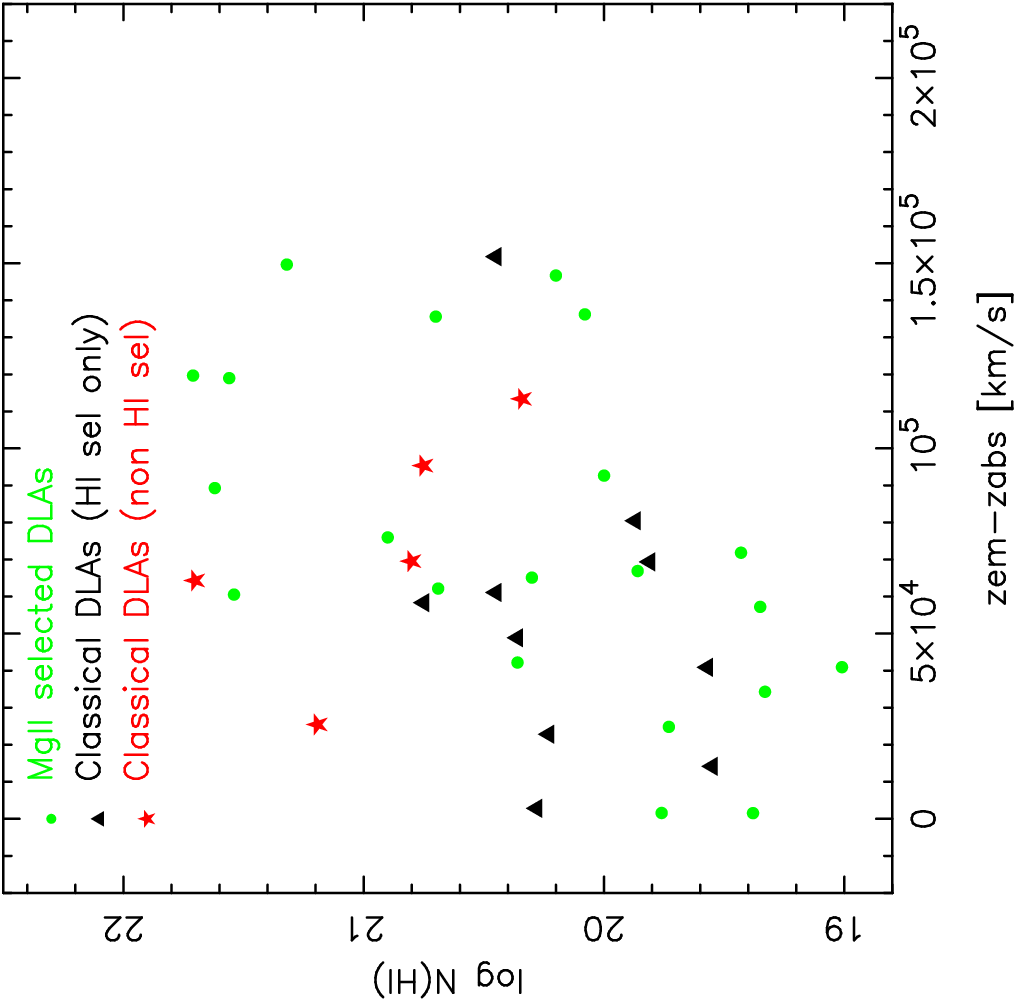}
\caption{\logh\ as a function of $z_{\rm em}-z_{\rm abs}$ in km/s
}
\label{f:vel}
\end{figure}

\subsection{Associated Systems and Evolutionary Effects}

Another scenario would have the absorbers associated with the quasar
rather than being truly intervening systems. In order to test this
hypothesis, we compute the velocity difference between the absorbers
and the quasars themselves on the basis that the former are actually
associated systems. The $z_{\rm em}-z_{\rm abs}$ in km/s can be
expressed as:

\begin{equation}
\frac{(z_{em}-z_{abs})\times c}{(1+z_{em})}
\end{equation}

where $c$ is the speed of light in km/s. The result
is shown in Figure~\ref{f:vel}. Interestingly, there seems to be a
hint of a correlation between \logh\ and $(z_{\rm em}-z_{\rm abs})$
for the \M-selected population, while this is not true for the
classical DLAs/sub-DLAs. In any case, we see no differences between
the two samples which could possibly explain their apparent
discrepancy in terms of \nhi\ distributions.

\begin{figure}
   \includegraphics[width=80mm,angle=-90]{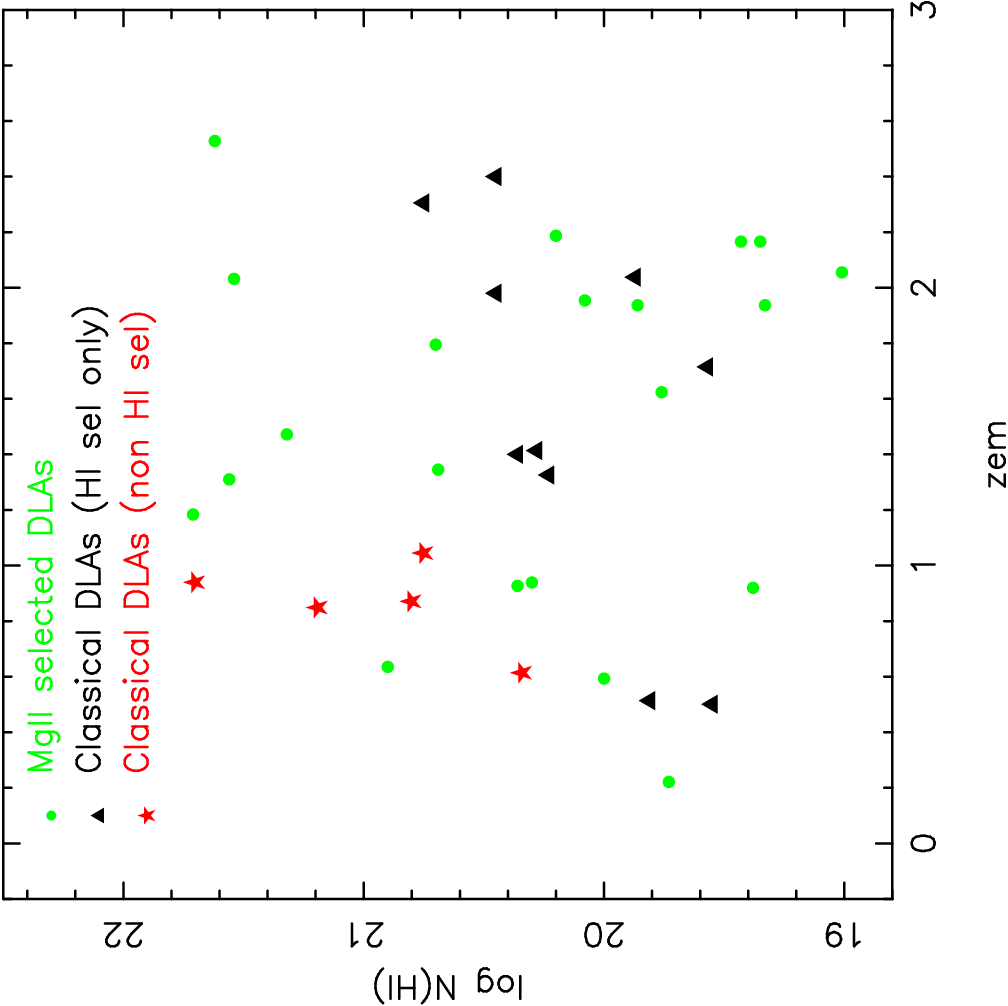}
\vspace{1.cm}		  
   \includegraphics[width=80mm,angle=-90]{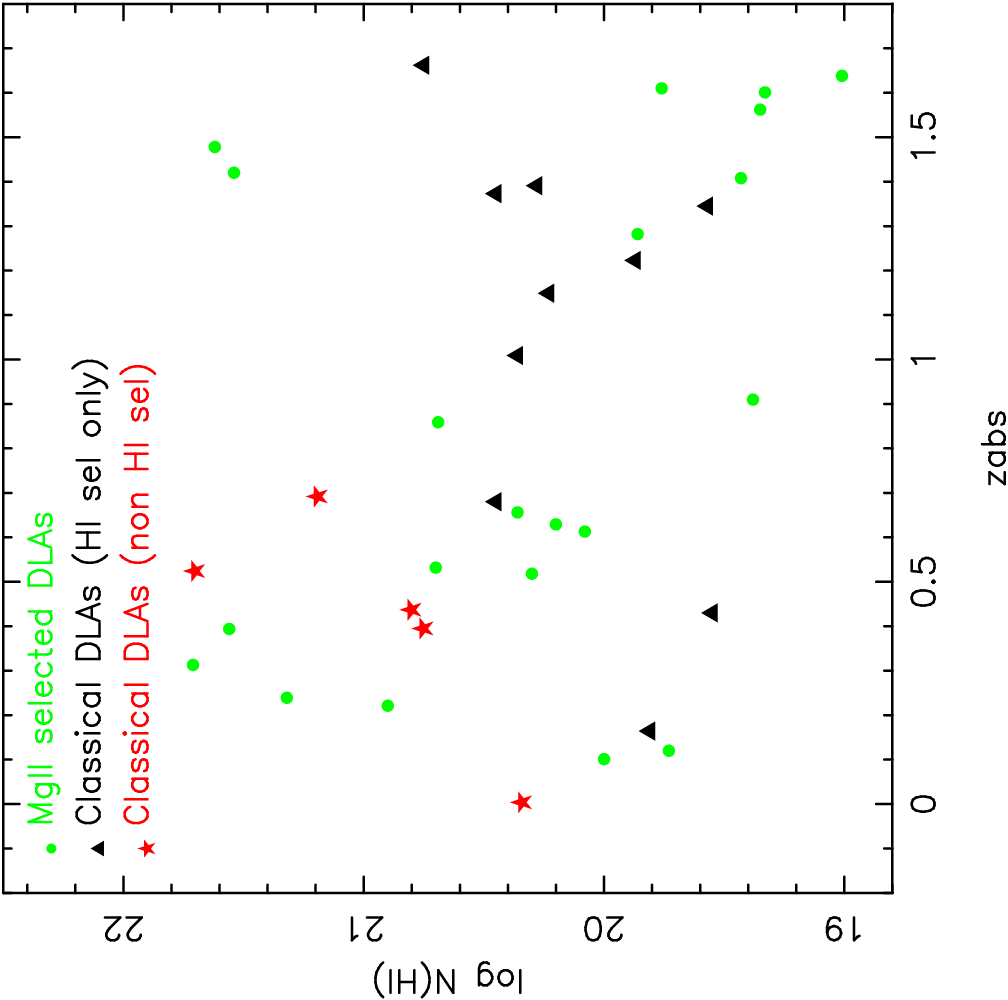}
\caption{{\it Top panel:} \logh\ as a function of quasar emission 
redshift ($z_{\rm em}$). {\it Bottom panel:} \logh\ as a function of
absorber redshift ($z_{\rm abs}$).}
\label{f:z}
\end{figure}

Finally, we test for potential evolutionary effects by analysing
\logh\ as a function of quasar emission redshift (top panel of
Figure~\ref{f:z}) and absorber redshift (bottom panel of
Figure~\ref{f:z}). Again, we find no obvious trend, and the two
populations of quasar absorbers appear to sample the same parameter
space.

\section{Discussion and Conclusion}

\hspace{1cm}

In this study, we compare the properties of the complete
samples of \M-selected absorbers and more ``classical'' DLAs at
$z_{\rm abs}<1.7$. Constructing well-defined samples has proven a
challenging task given the difficulties to build a homogeneous
\h-selected sample. Nevertheless, we find that the metal EW criterion 
used by Rao \& Turnshek (2000) to select DLAs traces well the
population of these systems down to the \logh$>$19.0. Moreover,
we find that the \M-selected sample presented has a marginally larger
fraction of absorbers with \logh$>$21.0 than seen otherwise at
low redshift. This property might in turn affect our estimates of
$\Omega_{\rm HI}$ which is dominated by the highest \h\ column
densities. 

We investigate the origin of the potential discrepancy and find that
\logh\ does {\it not} correlate in a significant way with either
metal equivalent widths or ratios of combination of equivalent widths,
quasar magnitude, quasar emission redshift, absorber redshift,
velocity distance from the absorber to the quasar, or position of the
absorber with respect to the quasar in physical space (a test for the
gravitational lensing hypothesis). We conclude that the marginal
discrepancy in \nhi\ distributions between the \M-selected ones and
those discovered with other techniques most likely arises from small
number statistics. 

We note that regardless of the physical reason for selecting
high-column density systems, these absorbers will directly affect
calculation of $\Omega_{\rm HI}$ at low redshift. In particular,
comparison of $\Omega_{\rm HI}$ calculations dominated by
\M-selected absorbers at $z_{\rm abs}<1.7$ with high-redshift systems
selected solely on \h\ might not be appropriate if the two methods are
not consistent. Furthermore, if there is a discrepancy between
classical and \M-selected absorption systems and if it holds at
high-redshift, it would imply that \h-selected surveys for
DLAs/sub-DLAs at z$>$2 have been missing a large fraction of the
neutral gas.

Since the \M\ method has proven an extremely efficient technique to
find low redshift absorbers, we anticipate that forthcoming
observations will help extend and refine the current
results. Therefore, larger samples are required in order to better
constrain $\Omega_{\rm HI}$ at $z<1.7$ and enable comparison with high
redshift results.

\section{Acknowledgments}
We thank Alberto Buzzoni for performing the TNG observations in
service mode, Marisa Girardi and Piercarlo Bonifacio for hints on blue
wavelength calibration, Jean-Claude Bouret for help with MAST and
Patrick Petitjean for comments on an earlier version of the
manuscript. We are grateful to the referee, Sandhya Rao, for openly
providing objective and expert suggestions which improved our
study. CP is supported by a Marie Curie fellowship. This work is
supported in part by the European Communities RTN network "The Physics
of the Intergalactic Medium". Based on observations made with the
Italian Telescopio Nazionale Galileo (TNG) operated on the island of
La Palma by the Centro Galileo Galilei of the INAF (Istituto Nazionale
di Astrofisica) at the Spanish Observatorio del Roque de los Muchachos
of the Instituto de Astrofisica de Canarias.

%

%

\bsp

\label{lastpage} 

\end{document}